\documentclass[twocolumn,pra,aps]{revtex4}
\usepackage{graphicx}

\begin{document}

\title{Leggett's inequality, the before-before experiment,
\\and free will on the part of Nature}

\author{Antoine Suarez}
\address{Center for Quantum Philosophy, P.O. Box 304, CH-8044 Zurich, Switzerland\\
suarez@leman.ch, www.quantumphil.org}

\date{January 20, 2008}

\begin{abstract} The before-before experiment demonstrates free will acting from outside space-time. The experimental violation of the Leggett's inequality supports the view that it is not appropriate to attempt to limit this freedom in Nature by forcing it to \emph{mimic} classical features.\\

\end{abstract}


\maketitle

Recently, several experiments testing Leggett's non-local hidden variables models have been presented \cite{grö,vs,az,cb}. Such models fulfill the so called Leggett's inequality, whereas quantum mechanics violates it \cite{le}. The experimental results show a violation of this inequality, and are in agreement with the predictions of quantum mechanics.

I have argued that Leggett's \emph{non-local realistic} models assuming time-ordered nonlocality, can be considered refuted by the before-before experiment \cite{as0708}. As far as one takes quantum nonlocality for granted, the before-before experiment demonstrates that Nature establishes non-local order without time, and rules out the view that an observable event (the effect) always originates from another observable event (the cause) occurring before in time \cite{szsg}.

The mathematical derivation of the Leggett's inequality as such does not require time-ordered nonlocality. This is particularly clear in \cite{cb} because the way the inequality is derived  avoids any description of the nonlocal links. By contrast, the inequality's derivation in \cite{grö} and \cite{az} uses mathematical terms expressing a dependency of an outcome on the other space-like separated one. However, one could consider that this dependency reflects a mere logical order, which is not determined by any inertial-frame, and therefore does not have any time-ordered counterpart in the physical reality.

Nevertheless, as soon as one embeds Leggett's inequality in a model supposed to describe the physical reality, one should take account of the before-before experiment using beam-splitters in motion \cite{szsg,as0708}, and characterize the non-local links: either  they are time-ordered (and hence timing-dependent \cite{as0708})or not.

If one embeds Leggett's inequality in a model assuming that one event can be considered the cause (occurring before in time), and the other the effect (occurring later in time), then the model is refuted by the before-before experiment, as said above.

Such an embedment occurs in the Leggett's models presented in \cite{grö,vs,az}:

In \cite{grö} and \cite{az} because one assumes the postulate of ``non-local realism'', according to which ``all measurement outcomes are determined by pre-existing properties of particles independent of the measurement''. This implies that one of the non-local correlated events precedes in time the other one \cite{as0708}.

In \cite{vs} because in the conclusion one explains the Leggett's model in the light of Bohmian mechanics, with ``a first particle to be measured'' and the other receiving thereafter the information: ``the particle that receives the communication is allowed to take this information into account to produce non-local correlations, but it is also required to produce outcomes that respect the marginals expected for the local parameters alone.'' This explanation clearly contradicts the conclusion in \cite{szsg}.

By contrast, in the more recent work \cite{cb} the authors avoid such an explanation and explicitly assume the result of the before-before experiment, stating that: ``nonlocal correlations happen from outside space-time, in the sense that there is no story in space-time that tells us how they happen.'' In this sense, the experimental results they present refute a Leggett's model that does not assume any time-order of the events \cite{cb}.

The experimental violation of the Bell's inequality led to the insight that Nature works out the quantum correlations faster than light. And the non disappearance of the correlations in relativistic experiments with before-before timing led to the insight that the quantum correlations originate from outside space-time \cite{as0705}. Can we derive a similar physical meaning from the experimental falsification of a Leggett's model without time-ordered nonlocality (like the model tested in \cite{cb})? I try to answer this question in the following.

First of all, I would like to stress that such a model can no longer be considered ``realistic'' in the sense of \cite{grö}.

Indeed, the before-before experiment demonstrates that Nature works out the quantum correlations in a nonlocal and non-deterministic way. This means that the measurement outcomes (for instance $A=+1$ and $B=-1$, in the experiment sketched in Figure 1) imply a true \emph{choice} on the part of Nature, and are not determined by pre-existing properties the particles carry independent of the act of measurement \cite{as0708,as0705}. Therefore it is no longer appropriate to think of trajectories and polarizations in the classical way. In particular it is confusing to imagine ``particles'' following continuous trajectories, since Nature's choices imply jumps. And it is confusing to assume that each photon is perfectly polarized in some direction, since the single local outcomes do not exclusively result from the properties of the polarized state. Thus, the postulate of ``realism'' as defined in \cite{grö} was already ruled out by the before-before experiment, and accordingly, Leggett's models compatible with this experiment should be considered ``non-realistic''.

The before-before experiment fully supports the orthodox view, and leads us to accept free will on the part of Nature additionally to the free will on the part of the experimenter (in accord with Anton Zeilinger's view of the ``two freedoms'' \cite{az05}).

Even after accepting this, one may be tempted to restrict Nature's freedom, forcing it to \emph{mimic} or \emph{simulate} certain classical features (trajectories, polarizations) that are not required by the statistical distributions imposed by the quantum formalism.

\begin{figure}[t]
\includegraphics[width=80 mm]{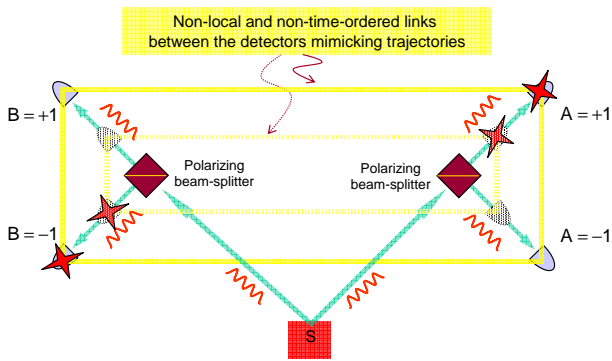}
\caption{Model assuming non-deterministic nonlocality and mimicking the concept of trajectory: A source S emits pairs of photons in a maximally entangled state. A polarization measurement gives either a result of $+1$ or $-1$ depending on whether the photon is transmitted or reflected by the corresponding polarizing beam-splitter. The yellow rods represent nonlocal links, which are not time-ordered. Supposed a coincident measurement gives the result: $A=+1$ and $B=-1$, the model assumes that the result would have been the same, if the measurement had been made with detectors less distant from the polarizing beam-splitters.}
\label{fig1}
\end{figure}

Consider the experiment in Figure 1, and suppose that a coincident measurement gives the result: $A=+1$ and $B=-1$. Suppose now counterfactually that the measurement had been made with detectors less distant from the polarizing beam-splitters. According to the Copenhagen or orthodox interpretation one cannot say that the result had been the same. Nature had very well had the freedom to produce another result. By contrast, if one wants Nature to mimic the ``realistic'' feature of trajectory, one will claim that the result would have been the same, as represented in Figure 1.

One can now conceive a model assuming nonlocality without time-order and forcing Nature to \emph{mimic} the feature of trajectory in the sense described above. To date there is no experimental falsification of such a model.

Extending this way of thinking, one may feel legitimate to force Nature to \emph{simulate} the ``realistic'' feature of polarization. Accordingly, the choices of the outcomes in the experiment of Figure 1 should fulfill the Malus law for each photon: results A, as though the photon A would meet the corresponding polarizing beam-splitter with a well-defined vector polarization \textbf{u}, and results B, as though the photon B would meet the corresponding polarizing beam-splitter with a well-defined vector polarization \textbf{v}. As we know, such a model bears outcomes fulfilling the Leggett's inequality, and therefore was ruled out by the experimental violation of this inequality \cite{grö,vs,az,cb}.

The experimental violation of the Leggett's inequality strengthens the view that it is not appropriate to restrict the freedom in Nature in the quantum phenomena by forcing it to imitate classical features. In this line of thinking it may be interesting to search for methods allowing us to test models that force Nature to \emph{mimic} the concept of trajectory, and in particular investigate the possibility of experimentally feasible schemes that claim to demonstrate nonlocality of a single particle \cite{vv}.

In conclusion, the before-before experiment is a key result to a proper understanding of nonlocal correlations. This experiment demonstrates agency from outside space-time, freedom in Nature (in accord with the orthodox view). The violation of the Leggett's inequality supports the view that it is misleading to attempt to limit this freedom through constraints stronger than the statistical distributions imposed by the quantum formalism.

\emph{Acknowledgments}: I am grateful to Cyril Branciard, Nicolas Brunner, Nicolas Gisin, and Valerio Scarani, for inspiring discussions.

\end{document}